\documentclass{iau} 
\usepackage{graphicx}

\title[Conditions in the young disk around L1527] 
{Unveiling the physical and chemical conditions in the young disk around L1527}

\author[M.L.R. van 't Hoff \etal]   
{M.L.R. van 't Hoff$^1$, J.J. Tobin$^2$, D. Harsono$^1$, E.F. van Dishoeck$^{1,3}$}

\affiliation{$^1$Leiden Observatory, Leiden University, PO box 9513,
NL-2300 RA, Leiden, the Netherlands \\ email: {\tt vthoff@strw.leidenuniv.nl} \\[\affilskip]
$^2$Homer L. Dodge Department of Physics and Astronomy, University of Oklahoma, 440 W. Brooks Street, Norman, OK 73019, USA \\[\affilskip]
$^3$Max-Planck-Institut f\"ur Extraterrestrische Physik, Giessenbachstrasse 1, 85748 Garching, Germany}

\pubyear{2017}
\volume{332}  
\jname{Astrochemistry VII -- Through the Cosmos from Galaxies to Planets}
\editors{M. Cunningham, T. Millar \& Y. Aikawa, eds.}

\begin{document}

\maketitle


\begin{abstract} Planets form in disks around young stars. The planet formation process may start when the protostar and disk are still deeply embedded within their infalling envelope. However, unlike more evolved protoplanetary disks, the physical and chemical structure of these young embedded disks are still poorly constrained. We have analyzed ALMA data for $^{13}$CO, C$^{18}$O and N$_2$D$^+$ to constrain the temperature structure, one of the critical unknowns, in the disk around L1527. The spatial distribution of $^{13}$CO and C$^{18}$O, together with the kinetic temperature derived from the optically thick $^{13}$CO emission and the non-detection of N$_2$D$^+$, suggest that this disk is warm enough ($\gtrsim$ 20~K) to prevent CO freeze-out.




\end{abstract}

       
\firstsection 
                     
\section{Introduction}

Disks around young stars are the birthplace of planets. The chemical structure of these disks, and thus of the material that will build up the planets, determines planet compositions. In addition, the disk physical structure influences the planet formation process. Therefore, evolved protoplanetary disks (Class II sources) have been intensively studied and are becoming well characterized both physically (e.g. \cite{Andrews2010}; \cite{Schwarz2016}) and chemically (e.g. \cite{Thi2004}; \cite{Dutrey2007}; \cite{Oberg2010}; \cite{Huang2017}). However, grain growth already starts when the protostellar system is still deeply embedded in its natal molecular cloud (\cite{Kwon2009}; \cite{Miotello2014}) and the HL Tau images support the idea that planet formation begins during this class 0/I phase (\cite{ALMA2015}). Thus, young embedded disks may reflect the true initial conditions for planet formation.

Although several embedded disks are now known (e.g. \cite{Tobin2012}; \cite{Murillo2013}; \cite{Harsono2014}; \cite{Aso2015}), and ALMA enables us to spatially resolve molecular emission from these young disks, their physical and chemical conditions remain unconstrained. One of the critical unknowns is the temperature structure, since this directly influences the volatile composition of the planet-forming material. In regions where the temperature exceeds $\sim$20~K, CO will be mainly present in the gas phase, while at lower temperatures CO is frozen out onto dust grains. In addition, this CO ice is the starting point for the formation of more complex molecules.  

 
\section{Probing the disk temperature structure with $^{13}$CO and N$_2$D$^+$}

\begin{figure}[t]
\begin{center}
 \includegraphics[width=\linewidth]{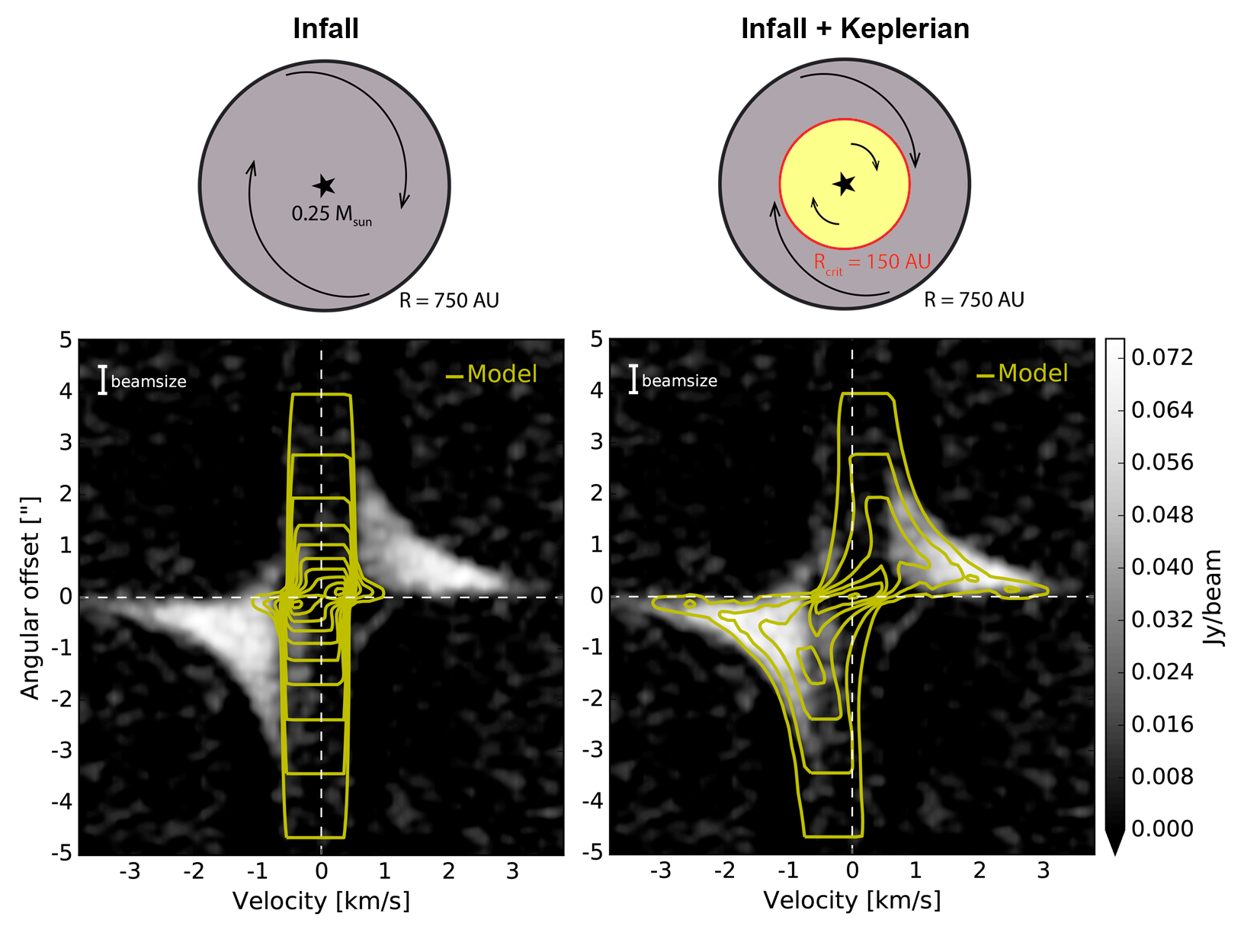} 
 \caption{Position-velocity diagram for $^{13}$CO observations (black-and-white scale) overlaid with contours for a thin disk model for a rotating infalling velocity profile (\textit{left panel}) and for a rotating infalling velocity profile with Keplerian rotation in the inner 150 AU (\textit{right panel}). The observations are taken with ALMA in Cycle 2 (PI: Koyamatsu).}
 \label{fig:pv}
\end{center}
\end{figure}

\begin{figure}[t]
\begin{center}
 \includegraphics[width=\linewidth]{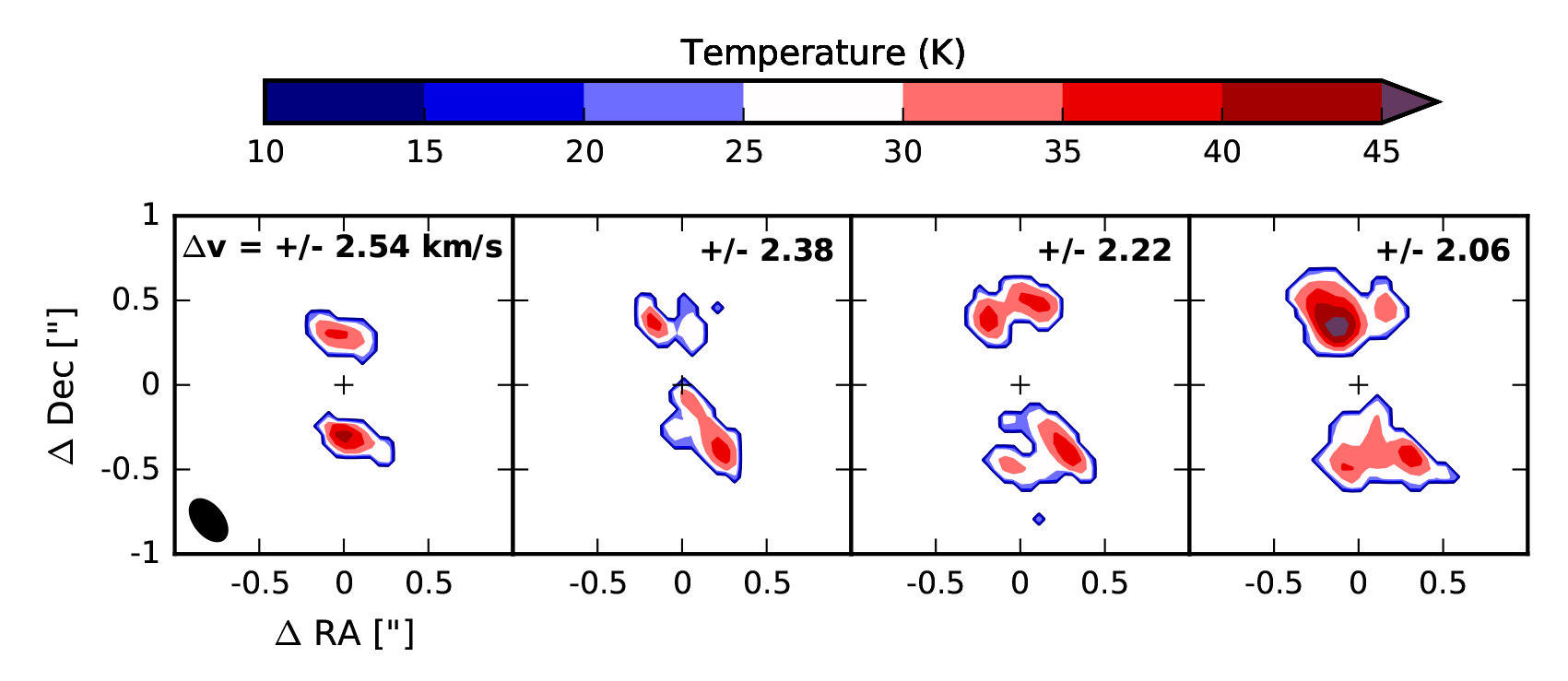} 
 \caption{Channelmaps for $^{13}$CO at the largest velocity offsets, that is, channels containing emission from the disk. Blue- and red-shifted channels with the same absolute velocity offset are shown in the same panel, and the velocity offset is shown in the top right corner. The beam is shown in the lower left corner of the first panel. CO freezes out at temperatures below $\sim$20~K. The observations are taken with ALMA in Cycle 2 (PI: Koyamatsu).}
 \label{fig:temp}
\end{center}
\end{figure}

We have analyzed our own and archival ALMA data (PIs: Tobin, Koyamatsu, Ohashi, Sakai) and used $^{13}$CO, C$^{18}$O and N$_2$D$^+$ observations to constrain the temperature in the embedded disk of L1527. This disk is particularly interesting because its almost edge-on configuration allows for direct probe of the vertical structure. Emission originating in the disk can be isolated from the envelope contribution based on velocity. Comparing the results from a thin disk model (\cite{Murillo2013}) with an infall velocity profile to the results from a model including Keplerian rotation in the inner region shows that the emission in the highest velocity channels is solely due to the disk (Fig.~\ref{fig:pv}). The spatial extent of the $^{13}$CO and C$^{18}$O $J$ = 2-1 emission in these channels suggest that CO is vertically present throughout the disk, including the disk midplane (Fig.~\ref{fig:temp}). In addition, the ratio of the $^{13}$CO and C$^{18}$O intensities shows that the $^{13}$CO emission is optically thick ($\tau > 3$), and thus traces the kinetic temperature of the gas. The derived temperatures are $\sim$30-40~K, above the CO freeze-out temperature of $\sim$20~K (Fig.~\ref{fig:temp}).

In contrast, N$_2$H$^+$ observations toward several protoplanetary disks have shown that the outer disk midplane becomes cold enough for CO to freeze out (\cite{Qi2013},\cite[ 2015]{Qi2015}). The N$_2$H$^+$ ion traces CO freeze-out because its main destructor is gas-phase CO (\cite{Aikawa2015}; \cite{vantHoff2017}). The deuterated form of N$_2$H$^+$, N$_2$D$^+$, is not detected in the L1527 disk, corroborating the observation that CO is present in the gas phase throughout the disk.

Altogether, these preliminary results are in agreement with physical models of embedded disks (\cite{Harsono2014}, \cite[ 2015]{Harsono2015}) and suggest that the young disk in L1527 is warm enough to prevent CO freeze-out.

\section*{Acknowledgments}

Astrochemistry in Leiden is supported by the European Union A-ERC grant 291141 CHEMPLAN, by the Netherlands Research School for Astronomy (NOVA) and by a Royal Netherlands Academy of Arts and Sciences (KNAW) professor prize. M.L.R.H acknowledges support from a Huygens fellowship from Leiden University.



\end{document}